\documentclass[doublecol]{epl2}
\usepackage{graphicx}
\usepackage{amsmath}
\input epsf
\newcommand{\be}{\begin{equation}}          % redefines \begin{equation}
\newcommand{\ee}[1]{\label{#1} \end{equation}}  % redefines \end{equation}
\newcommand{\bee}{\begin{eqnarray}}         % redefines \begin{eqnarray}
\newcommand{\eee}{\end{eqnarray}}            % redefines \end{eqnarray}
\newcommand{\avg}[1]{\left\langle#1\right\rangle}
\newcommand{\braces}[1]{\left\lbrace#1\right\rbrace}

\title{Diverse routes to oscillation death in a coupled oscillator system}
\author{Jos\'e~J.~Su\'arez-Vargas\inst{1} \and Jorge~A.~Gonz\'alez\inst{1}
\and Aneta~Stefanovska\inst{2} \and Peter~V.~E.~McClintock\inst{2} }
\shortauthor{Su\'arez-Vargas, Gonz\'alez, Stefanovska, McClintock}
\institute{
\inst{1} Physics Center, Venezuelan Institute for Scientific Research,
Caracas 1020-A, Venezuela.\\
\inst{2} Physics Department, Lancaster University, Bailrigg, LA1 4YB, Lancaster, UK.}
\date{\today}

\abstract{We study oscillation death (OD) in a well-known
coupled-oscillator system that has been used to model cardiovascular
phenomena. We derive exact analytic conditions that allow the prediction
of OD through the two known bifurcation routes, in the same model, and
for different numbers of coupled oscillators. Our exact analytic results
enable us to generalize OD as a multiparameter-sensitive phenomenon. It
can be induced, not only by changes in couplings, but also by changes in
the oscillator frequencies or amplitudes. We observe
synchronization transitions as a function of coupling and confirm
the robustness of the phenomena in the presence of noise. Numerical
and analogue simulations are in good agreement with the theory.}

\begin{document}
\maketitle
%\section{Introduction}

Coupled oscillator systems exhibit a variety of phenomena
relevant to physics, biology, and other branches of science and
technology. Here, we study \textit{oscillation death} (OD)
\cite{ODAD}, a form of synchronization \cite{Pikovsky:01} in
which the oscillators interact in such a way as to quench each
other's oscillations
\cite{Bar-Eli:85,Crowley:89,Aronson:90,Zhu:08}. This intriguing
phenomenon was noted in the 19th century by Rayleigh
\cite{Pikovsky:01}, who found that adjacent organ pipes of the
same pitch can reduce each other to silence. Since then, OD has
been studied in diverse applications including oceanography
\cite{Gallego:01}, chemical engineering \cite{Zhai:04},
solid-state lasers \cite{Wei:07} and a variety of other
experimental systems
\cite{Crowley:89,Ozden:04,Herrero:00,Reddy:00}. OD is known to
occur via two distinct bifurcation mechanisms: (i) Hopf
bifurcation, where the coupling induces stability at the origin
of the phase space, thus collapsing the orbits to zero, which
can happen only if the oscillators are sufficiently different
\cite{Aronson:90,Ermentrout:90a,Ermentrout:90b,Mirollo:90,Matthews:90}
(or for identical oscillators if there are delays
\cite{Reddy:98,Atay:03} in the coupling); or (ii) for
non-identical oscillators, saddle-node bifurcation
\cite{Crowley:89} in which new fixed points appear on/near the
coupled limit cycles, annihilating the periodic orbits.
Recently, Karnatak \textit{et al}. \cite{Karnatak:07} were able
to produce OD in two identical coupled oscillators through the
saddle-node route, using dissimilar non-delayed coupling.

In this Letter, we show that a coupled-oscillator system, which
has been used extensively in modeling coupled rhythmic
processes in mathematics, physics and biology \cite{Glass:88}
can undergo OD via both bifurcation routes. We obtain exact
analytic conditions for OD, and compare the theory with
numerical simulations and analogue electronic experiments. We
thus generalize OD as a phenomenon that occurs, not only
through coupling-increased dissipativity \cite{Pikovsky:01},
but also when a measure of dispersion among the parameters of
the coupled system is exceeded \cite{De_Monte:03}. We also show
that, near the onset of death, the coupled system alternates
between periodic, quasiperiodic and even chaotic behavior,
reflecting the complex temporal variability observed in real
biological systems \cite{Ivanov:99,Volkov:94}.

Our model is a set of five coupled oscillators that
successfully reproduces many phenomena seen in the
cardiovascular system (CVS), e.g.\ modulation
\cite{Stefanovska:99a} and synchronization
\cite{Stefanovska:01a}. Each oscillator has its own
characteristic frequency and amplitude \cite{Stefanovska:07}
(see Table 1) and emulates a particular physiological function.
Heart and respiration are obvious physiological
processes; the myogenic oscillation is related to the intrinsic
self-regulatory activity of the smooth muscle tissue in the
walls of the blood vessels; the neurogenic oscillation is
associated with the neural control by the central nervous
system; and the nitric oxide related endothelial oscillation
is associated with metabolic activity mediated by
the endothelial tissue that lines the whole CVS \cite{foot1}.

The basic unit,
\begin{equation}
\label{Poinc:eq1}
\begin{array}{c c l}
\dot{x}_i&=&-x_iq_i-y_i\omega_i + P_i(\textbf{X},\mathbf{Y}) \\
\dot{y}_i&=&-y_iq_i+x_i\omega_i + Q_i(\textbf{X},\mathbf{Y}),
\end{array}
\end{equation}
is the Poincar\'e oscillator \cite{Glass:88} with:
$i=1,\ldots,m$, $(m=5)$;
$q_i=\alpha\left(\sqrt{(x_i^2+y_i^2)}-a_i\right)$; $a_i$,
$\omega_i=2\pi f_i$ and $\alpha$ are constants that represent amplitude, frequency and dissipation rate respectively. \textbf{X} and
\textbf{Y} $\in R^m$, and $P_i, Q_i:R^m \rightarrow R$ are
scalar coupling functions. We suppose that the inter-oscillator
interactions can be approximated by a mean field, so that
$P_i(\mathbf{X},\mathbf{Y})=\frac{1}{m}\displaystyle\sum{ x_j
}$, and $Q_i(\mathbf{X},\mathbf{Y})=0$.

\begin{table}
\caption{Typical values of frequency (Hz) and relative
amplitude (arbitrary units) of physiological rhythms in humans,
as measured in blood flow by laser-Doppler flowmetry and
analysed by wavelet transform. The characteristic frequencies
vary around the values indicated; the amplitudes are only estimated based
on measurements \cite{Stefanovska:99a,Stefanovska:07}.}
\label{empir_values}
\begin{center}
\begin{tabular}{lcr}
\textbf{Activity}  & \textbf{Frequency} & \textbf{Amplitude}\\
Heart & 1.1 Hz & 0.5 a.u.\\
Respiration & 0.3 Hz & 1.0 a.u.\\
Myogenic & 0.1 Hz & 1.0 a.u.\\
Neurogenic & 0.04 Hz & 1.0 a.u.\\
Endothelial & 0.01 Hz & 0.5 a.u.\\
\end{tabular}
\end{center}
\end{table}

We now review briefly some basic concepts, assuming an autonomous
$n$--dimensional dynamical system:
\begin{equation}
\begin{array}{c c l}
\label{dyn_syst:eq1}
\dot{\textbf{Z}}&=&\textbf{F}(\textbf{Z}),
\end{array}
\end{equation}

\noindent where $\textbf{Z} \in R^n$ and $\textbf{F}:R^n
\rightarrow R^n$ is a general nonlinear vector function. All the
points, $\textbf{Z}^*$, in phase space satisfying the equation
$\textbf{F}(\textbf{Z}^*)=\textbf{0}$, are called fixed points of
the dynamical system (\ref{dyn_syst:eq1}). Their stability and
quality can be determined by the eigenvalues of the Jacobi matrix,
provided their real parts are nonzero:
\begin{equation}
\label{Detet:Jacobi}
\det\braces{\frac{\partial \textbf{F}(\textbf{Z})}{\partial \textbf{Z}} \vert_{\textbf{Z}=\textbf{Z}^*}-\lambda \textbf{I} }=0
\end{equation}

%\subsection{Two Oscillators}
\noindent We consider first the case of two coupled oscillators,
modelling e.g.\ the cardio-respiratory interactions,
\begin{equation}
\label{Two_Oscill}
\begin{array}{c c l}
\dot{x_1}&=&-x_1\alpha(\sqrt{x_1^2+y_1^2}-a_1)-y_1\omega_1+
\epsilon(x_1+x_2),\\
\dot{y_1}&=&-y_1\alpha(\sqrt{x_1^2+y_1^2}-a_1)+x_1\omega_1,\\

\dot{x_2}&=&-x_2\alpha(\sqrt{x_2^2+y_2^2}-a_2)-y_2\omega_2 +
\epsilon(x_1+x_2),\\
\dot{y_2}&=&-y_2\alpha(\sqrt{x_2^2+y_2^2}-a_2)+x_2\omega_2.
\end{array}
\end{equation}

\noindent Here the origin is always a fixed point. For $\epsilon =
0$, this point is unstable (focus or node).  Besides this fixed
point, there is a stable limit cycle, corresponding to autonomous oscillations.
Eq.~(\ref{Detet:Jacobi}) for the point $\textbf{Z}^*=\textbf{0}$
and $\epsilon \neq 0$ in the dynamical system ~(\ref{Two_Oscill})
will be:
\begin{equation}
\label{Detet:Jacobi2}
\left|\begin{array}{c c c c}
(\alpha a_1+\epsilon)-\lambda&-\omega_1&\epsilon&0\\
\omega_1&\alpha a_1-\lambda&0&0\\
\epsilon&0&(\alpha a_2+\epsilon)-\lambda&-\omega_2\\
0&0&\omega_2&\alpha a_2-\lambda
\end{array}\right|=0.
\end{equation}

\noindent From (\ref{Detet:Jacobi2}) we determine that the
origin can become a stable fixed point if
\begin{equation}
\label{condition:eq1}
\begin{array}{c}
2\alpha a_1+\epsilon < 0,\\
2\alpha a_2+\epsilon < 0,
\end{array}
\end{equation}

\noindent in which case the limit cycle no longer exists. Thus,
if $a_2 > a_1$, then $\epsilon < - 2 \alpha a_2$ is a
sufficient condition for OD. This is the Hopf bifurcation route
to OD, and it occurs only for negative $\epsilon$.
\begin{figure}
\begin{center}
\begin{tabular}{c}
\includegraphics*[width=8.3cm]{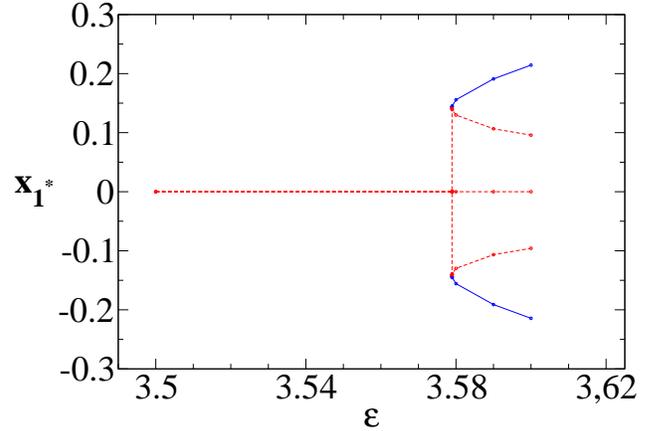}
\end{tabular}
\end{center}
\caption[Bifurcation]{\label{bif_diag}(Color online) Saddle-node bifurcation diagram showing the appearance of new equilibria and their
stability for coupled system (\ref{Two_Oscill}). The dashed
(red) line corresponds to unstable fixed points, while the
continuous (blue) line depicts stable ones.}
\end{figure}

For $\epsilon > 0$, there is a critical value $\epsilon_c$ such
that, for $\epsilon > \epsilon_c$, four new fixed points
appear: two unstable and two asymptotically stable. In order to
make numerical estimations from our model we make use of the
extensive earlier research on the CV system
\cite{Stefanovska:99a,Stefanovska:07} embodied in the
frequencies and amplitude relationships summarised in Table 1,
so that we take $a_1/a_5\approx 1$, $a_1/a_i\approx 0.5$, for
$i=\braces{2,3,4}$.

Fig.~\ref{bif_diag} shows the bifurcation diagram for $x_1$,
calculated for $a_1=0.5$, $a_2=1$, $f_1 = 1.1$, $f_2 = 0.3 $
and $\alpha=1$. Note that for this range of $\epsilon>0$ the
origin is always an unstable fixed point. The end of the
oscillations is marked by the appearance of the new fixed
points at $\epsilon\sim 3.578$. They are obtained from the set
of algebraic equations:

\begin{equation}
\label{Two_Oscill:eq2}
\begin{array}{l c c}
-x_1\alpha(\sqrt{x_1^2+y_1^2}-a_1)-y_1\omega_1 +
\epsilon(x_1+x_2)&=&0,\\
-y_1\alpha(\sqrt{x_1^2+y_1^2}-a_1)+x_1\omega_1&=&0,\\
\\
-x_2\alpha(\sqrt{x_2^2+y_2^2}-a_2)-y_2\omega_2 +
\epsilon(x_1+x_2)&=&0,\\
-y_2\alpha(\sqrt{x_2^2+y_2^2}-a_2)+x_2\omega_2&=&0.
\end{array}
\end{equation}

\noindent After some algebra, we obtain the relations:
\begin{equation}
\label{condition:eq2}
\begin{array}{c c c}
\omega_1(x_1^2+y_1^2)&=&\epsilon y_1(x_1+x_2),\\
\omega_2(x_2^2+y_2^2)&=&\epsilon y_2(x_1+x_2).
\end{array}
\end{equation}

\noindent Analysis of Eq.~(\ref{condition:eq2}) combined with
Eq.~(\ref{Two_Oscill:eq2}) shows that, for $\epsilon \ll
2\omega_1$ and $\epsilon \ll 2\omega_2$, the only fixed point is
$(0, 0, 0,0)$. For sufficiently large values of $\epsilon$, we
obtain four new fixed points. The fixed points can appear only in
pairs of stable--unstable points. This is the saddle-node
bifurcation route to OD, which occurs only for positive $\epsilon$.

We now define  $x_1 = r_1 \cos{(\Phi_1)}$, $y_1 = r_1
\sin{(\Phi_1)}$, $x_2 = r_2 \cos{(\Phi_2)}$, $y_2 =
r_2\sin{(\Phi_2)}$, where $r_1^2 = x_1^2 + y_1^2$, $r_2^2 = x_2^2 +
y_2^2$. As in our model $\omega_1\gg \omega_2$, $a_2 > a_1$, which
implies $(x_1^0)^2 \ll (x_2^0)^2$. From Eq.~(\ref{condition:eq2}) we
obtain the equality:
\begin{equation}
\label{condition:eq3} \omega_2 = \epsilon
\cos{(\Phi_2)}\sin{(\Phi_2)}.
\end{equation}

\noindent This is equivalent to the equation:

\begin{equation}
\label{condition:eq4}
\displaystyle\frac{2\omega_2}{\epsilon}=\sin{(2\Phi_2)}.
\end{equation}

\noindent As $\sin{(2\Phi_2)}\le 1$, this means that we have new
solutions to the algebraic equations (\ref{Two_Oscill:eq2}) only
when
\begin{equation}
\label{condition:eq5}
\epsilon > 2\omega_2.
\end{equation}

\noindent Eq.~(\ref{condition:eq5}) provides a simple and
understandable analytic estimate of the critical value for the
bifurcation.

\begin{figure}
\begin{center}
\begin{tabular}{c}
\includegraphics*[width=8.3cm]{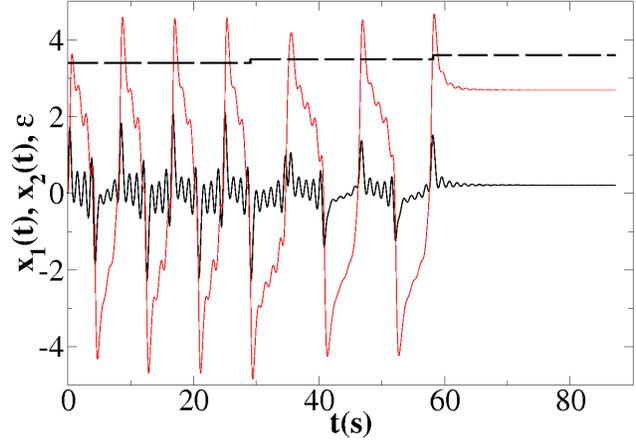}\\
\includegraphics*[width=8.3cm]{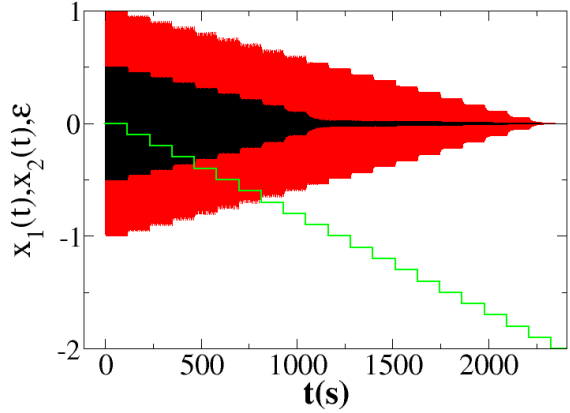}
\end{tabular}
\end{center}
\caption{\label{x1_x2}(Color online) Top: Time series of $x_1(t)$
(bold curve) and $x_2(t)$ from numerical simulations of the
system (\ref{Two_Oscill}), showing saddle-node OD. Parameter
values were the same as in Fig.~\ref{bif_diag}. The dashed line
shows the step--wise variation of the coupling constant
$\epsilon$ in a positive range. Note that after OD has occurred
$\displaystyle \lim_{ t\rightarrow
\infty}\braces{x_1(t),x_2(t)}> 0$. Bottom: Time series
of $x_1(t)$ and $x_2(t)$ from numerical simulations of the
system (\ref{Two_Oscill}), showing supercritical Hopf
bifurcation OD. Parameter values were the same as in
Fig.~\ref{bif_diag}. The diagonal stepped line starting at the
origin indicates how $\epsilon$ was varied in the negative
range. In this case $ \displaystyle\lim_{ t\rightarrow \infty}
\braces{x_1(t),x_2(t)} = 0$, after OD has occurred. Note the
difference in abscissa timescales between the top and bottom
parts of the figure.}
\end{figure}

Fig.~\ref{x1_x2} (top) illustrates a numerical simulation of the coupled
oscillators (\ref{Two_Oscill}) near criticality, showing how the
oscillations evolve as $\epsilon$ is increased by steps of 0.1. We
have established that, for the chosen parameter values, the
oscillations die when $\epsilon_c\approx 3.578$, gratifyingly %\gtrsim
close to the $\epsilon_c\approx 4\pi f_2 = 3.77$ predicted by
(\ref{condition:eq5}). The difference is attributable to the
approximations made in deriving Eq.~(\ref{condition:eq5}). Using
the same values of parameters, we also performed numerical
simulations for negative $\epsilon$, Fig.~\ref{x1_x2} (bottom), and found that OD occurs through supercritical Hopf bifurcation when $\epsilon\sim -1.88$, in agreement with the theoretical prediction (\ref{condition:eq1}).

We have also used the above mathematical tools to investigate
the full model \cite{Stefanovska:99a} with five coupled oscillators $(i=1,\ldots,5)$.
Solutions of the algebraic equations
\begin{equation}
\label{five_algeb:eq1}
\begin{array}{c c l}
0&=&-x_iq_i-y_i\omega_i + \epsilon \displaystyle \sum_{j=1}^5{x_j} \\
0&=&-y_iq_i+x_i\omega_i ,
\end{array}
\end{equation}
correspond to the fixed points of (\ref{Poinc:eq1}) for five
coupled oscillators. After some calculations we get the equations
\begin{equation}
\label{five_algeb:eq2}
\begin{array}{c c c}
\omega_i(x_i^2+y_i^2)&=&\epsilon y_i\sum{ x_j}.
\end{array}
\end{equation}

\noindent We have found that, for $\epsilon > \epsilon_c$, where
$\epsilon_c$ is some critical value, the system possesses four
additional fixed points (two stable and two unstable).

Because $\omega_i=2\pi f_i$, and using the empirical physiological values
(table~\ref{empir_values}) for the $f_i$'s and $a_i$'s mentioned above, we can use the inequalities $\omega_5 <\omega_4 <\omega_3 <\omega_2 <\omega_1$, and $a^2_1=a^2_5 < a^2_2=a^2_3=a^2_4$, in order to obtain an analytic expression for $\epsilon_c$. The resulting approximate equations lead to the following relations for the systems' fixed points: $\sum{x_j}\approx \omega_1a_1$; $x_1^o\approx y_4^o \approx y_5^o \approx 0$; $x_3^o\approx x_4^o \approx x_5^o \approx x_*$, where $x_*=\frac{a_5\alpha+\sqrt{a_5^2\alpha^2+4\omega_1a_1\alpha}
}{2\alpha } $; $y_1\approx a_1$; $y_2^o=\frac{(a_1\omega_1
-a_2^2\alpha)+\sqrt{(\omega_1 a_1-\alpha
a_2^2)^2+4a_2^3\alpha\omega_2} }{2\omega_2}$

The new simplified equation (\ref{five_algeb:eq2}), with $i=2$,
is
\begin{equation}
\label{five_algeb:eq3}
\begin{array}{c c c}
\omega_2((x_2^o)^2+(y_2^o)^2)&=&\epsilon y_2^ox_2^o +\epsilon 3x_* y_2^o.
\end{array}
\end{equation}
The new stable fixed points are possible when
\begin{equation}
\label{five_algeb:eq4}
\begin{array}{c c c}
y_2^o\epsilon^2 + 12x_*\omega_2\epsilon - 4\omega_2^2 y_2^o &>&0.
\end{array}
\end{equation}
The asymptotically stable equilibrium points are the new attractors of the dynamical system.

\begin{figure}[!t]
\begin{center}
\begin{tabular}{c}
\includegraphics*[width=8.3cm]{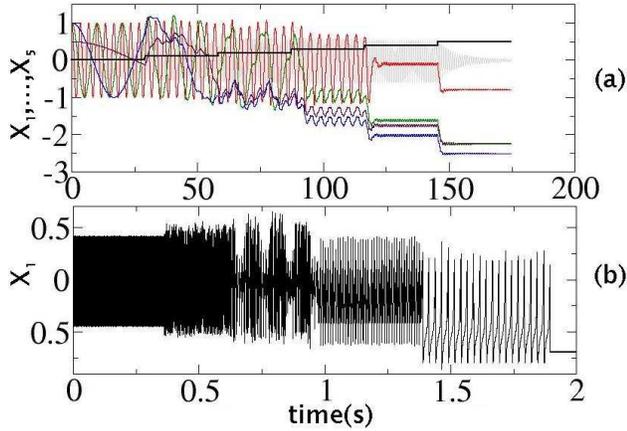}
\end{tabular}
\end{center}
\caption{(Color online) Simulations of the coupled system
(\ref{Poinc:eq1}) using five oscillators, as $\epsilon$ is
increased in steps of 0.1. Parameter values are based on
physiology (see text). (a) Numerical simulation showing the
outputs from all five oscillators, exhibiting OD for $\epsilon
\approx 0.47$. (b) Analogue electronic simulation, showing the %\gtrsim
output from the first oscillator $x_1$.} \label{Five_Oscill_Numerical}
\end{figure}

Using physiologically estimated values for the parameters (table~\ref{empir_values}) in
(\ref{five_algeb:eq4}) we obtain that $\epsilon_c\approx 0.468$.
This result is in agreement with the numerical simulations shown
on Fig.~\ref{Five_Oscill_Numerical}(a), where there is OD for
$\epsilon\approx 0.47$. %\grtsin

We have also modeled the dynamics of the five-oscillator system
for the given parameter values using analogue electronic circuits \cite{Luchinsky:98}.
% Fig.\ ~\ref{poinc_circ} shows one equivalent electronic circuit of Eq.\ ~(\ref{Poinc:eq1}).
The result of Fig.~\ref{Five_Oscill_Numerical}(b) shows the experimental observation of OD in good qualitative agreement with both the theory and numerical simulations.

\begin{figure}[!t]
\begin{center}
\begin{tabular}{c}
\includegraphics*[width=8.3cm]{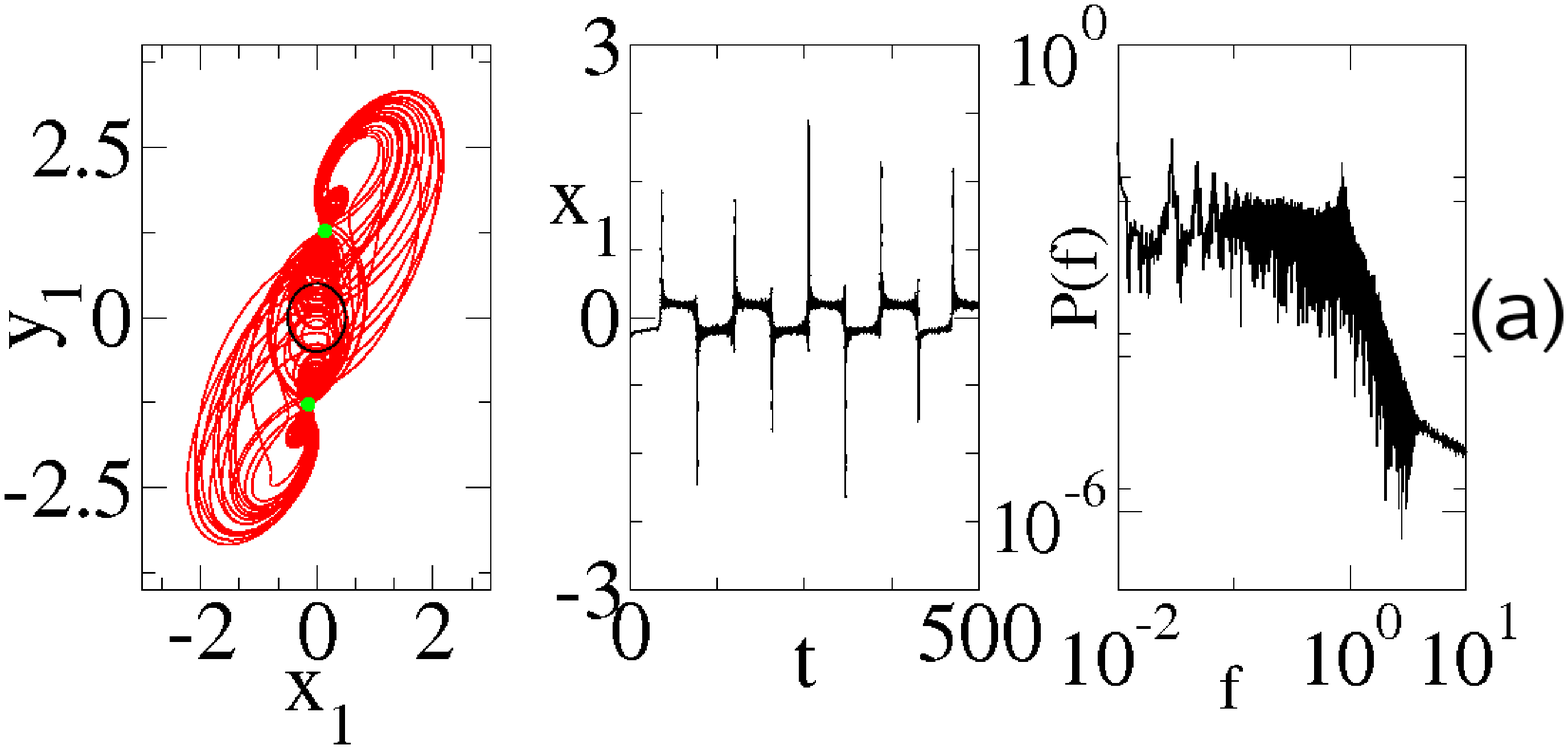}\\
\includegraphics[width=8.3cm,clip]{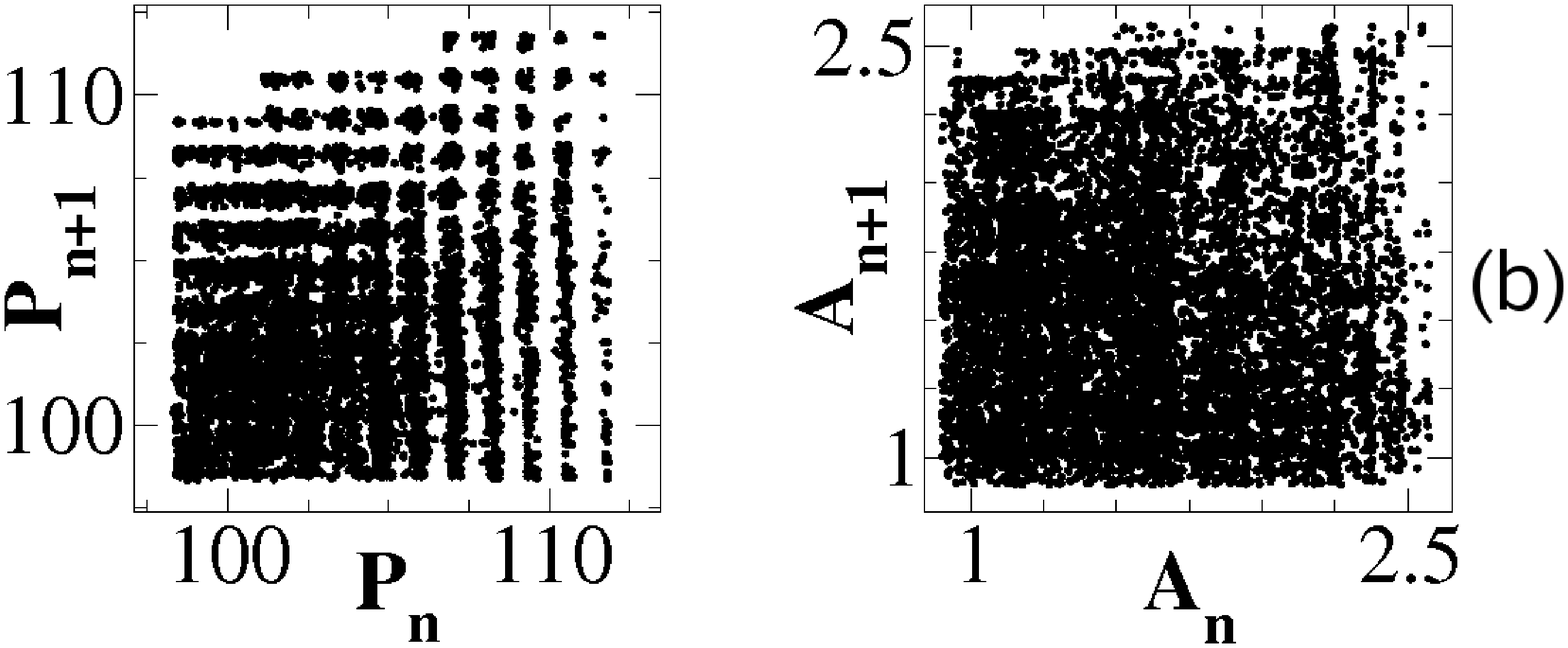}
\end{tabular}
\end{center}
\caption{(Color online) Behavior of the coupled system (\ref{Two_Oscill}) for $\epsilon=3.578$, very close to death. Row (a): Attractor in $x_1$-$y_1$ space, time series $x_1(t)$ and its Fourier transform. Row (b): First-return maps of the instantaneous period and amplitude signals.}
\label{AttractorsVsKappa}
\end{figure}

We have found that, just before the onset of death, the
2-oscillator system (\ref{Two_Oscill}) exhibits highly complex,
quasiperiodic and chaotic, behavior. Fig.~\ref{AttractorsVsKappa}(a)
plots the $x_1$-$y_1$ phase diagram when
$\epsilon=3.578\approx \epsilon_c$. For comparison the circle %\lesssim
in the background shows the limit cycle when $\epsilon=0$, and
the two full (green) spots signal the place where the new fixed
points will arise. The attractor's structure, the seemingly
random amplitudes of the time-series, the lengthening of the
period with a square-root scaling, and the power-law-like
spectrum, indicate that the system enters a chaotic regime just
before it dies. Fig.~\ref{AttractorsVsKappa}(b) shows the
first-return maps (FRM) of the discrete instantaneous period
and amplitude signals obtained by sampling $x_1(t)$ at the
peaks. We observe strong period and amplitude variability in
the FRM, with the period showing a multimodal distribution. The
amplitude of peaks has a more uniform distribution showing
fewer preferred values, and the amplitude difference between
successive peaks is distributed normally. In terms of the
cardiovascular analogue this complex behavior might be seen as
a predictor of imminent death of the cardio-respiratory coupled
oscillations. Moreover, multimodal period distributions are
typical of the time variability observed in other biological
systems, e.g.\ the intermitotic time of human skeletal cells,
and moments of change in the rotation direction of flagella
\cite{Volkov:94}. It is evident that, at least in this kind of
coupled biological system, the variability of the oscillations
increases near the onset of global bifurcations.

Thus, OD in (\ref{Poinc:eq1}) can occur via both of the known routes.
When $\epsilon$ is negative and below a critical value, the origin
becomes asymptotically stable and the oscillations die at almost
constant frequency, i.e.\ the Hopf bifurcation scenario. In the
second mechanism, OD occurs when new fixed points appear on the
former attractor after $\epsilon$ has surpassed a positive threshold:
the amplitude remains almost constant while the frequency decreases
with a square-root scaling in a saddle-node bifurcation.

When a system arrives in the OD regime it lies
quiescent. Although apparently trivial, this state results from
diverse complex interactions between the coupled elements that
form the system, as we show in this Letter. Thus OD can be seen
as another kind of complex collective motion, much in the same
way as synchronization arises in complex coupled systems as a
self-organized dynamics.

In order to complete the analysis of system (\ref{Poinc:eq1})
in this context we investigate the relationship between
synchronization and OD by measuring the synchronization index
\cite{Pikovsky:01}, defined as:
$\gamma_{1,1}=\avg{\cos{\Psi_{1,1}}}^2+\avg{\sin{\Psi_{1,1}}}^2$,
where $\Psi_{1,1}$ is the relative phase difference between the
oscillators. This index measures quantitatively the strength of
1:1 synchronization.

\begin{figure}[t!]
\begin{center}
\begin{tabular}{c}
\includegraphics*[width=8.3cm]{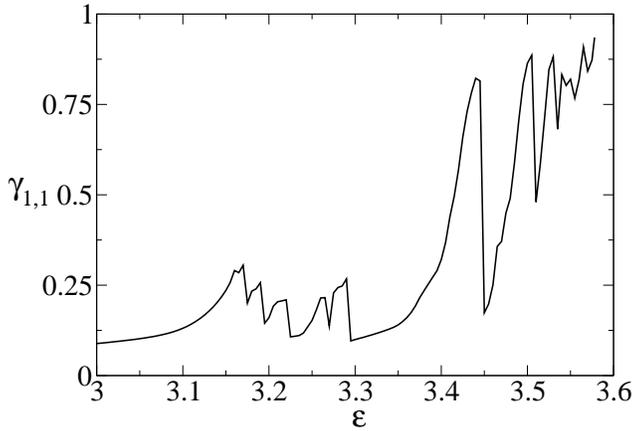}
\end{tabular}
\end{center}
\caption{\label{synch_fig}Synchronization index $\gamma_{1,1}$
as a function of the coupling strength $\epsilon$ for the
system (\ref{Poinc:eq1}).}
\end{figure}

Fig.~\ref{synch_fig} shows the calculated values of
$\gamma_{1,1}$ as a function of $\epsilon$ for the coupled
system (\ref{Poinc:eq1}). Initially $\gamma_{1,1}$ is very
small for low values of $\epsilon$ and it increases
monotonically in a quasi-exponential fashion as $\epsilon$ is
increased. However when $\epsilon$ reaches a value near 3.15,
$\gamma_{1,1}$ no longer grows monotonically, but has
alternating epochs of growth and decrease, corresponding to
topological changes in the structure of the attractors of both oscillators that lead to complex transitions in the 1:1 synchronization state. These
results suggest that the transient epochs of cardio-respiratory
synchronization seen (for higher synchronization ratios) in
many studies of resting humans, both awake
\cite{Schaefer:99,Lotric:00a,Kenwright:08} and asleep
\cite{Bartsch:07}, may arise in part from changes in coupling,
as well as from drifts in the natural frequencies and other
parameters.

Finally, in order to explore the robustness of our results in a
more realistic framework, we performed simulations of system
(\ref{Poinc:eq1}) including the presence of stochastic forces,
since they can lead to spurious detection of complex phenomena
\cite{Xu:06}. Our stochastic model is:

\begin{equation}
\label{Poinc:noise}
\begin{array}{c c l}
\dot{x}_i&=&-x_iq_i-y_i\omega_i +\epsilon(x_1+x_2)+D\xi_x\\
\dot{y}_i&=&-y_iq_i+x_i\omega_i,
\end{array}
\end{equation}
where $\xi_x$ corresponds to white Gaussian noise with zero
mean and variance $D^2$.

\begin{figure}
\begin{center}
\begin{tabular}{c}
\includegraphics*[width=8.3cm]{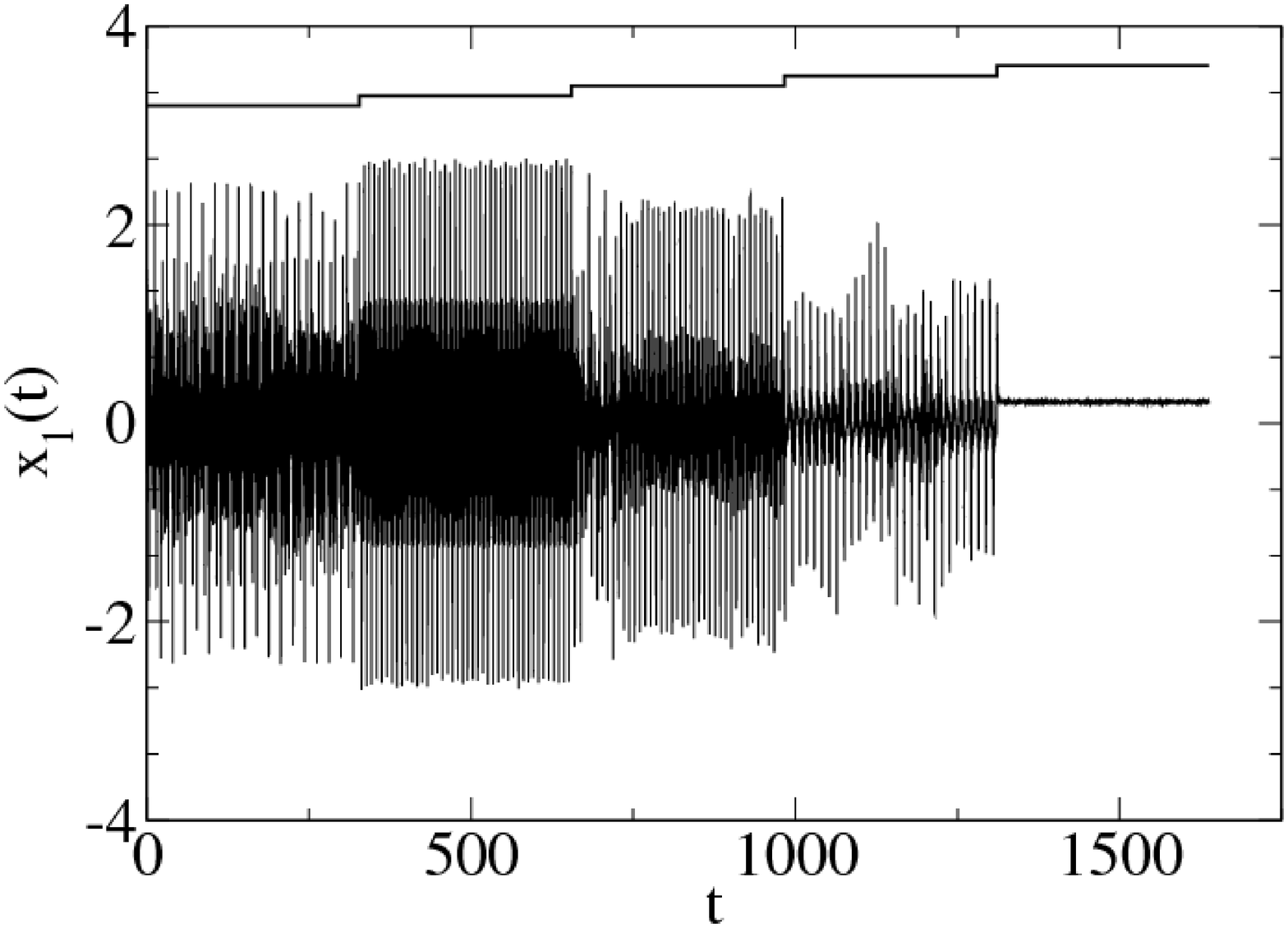}\\
\includegraphics*[width=8.3cm]{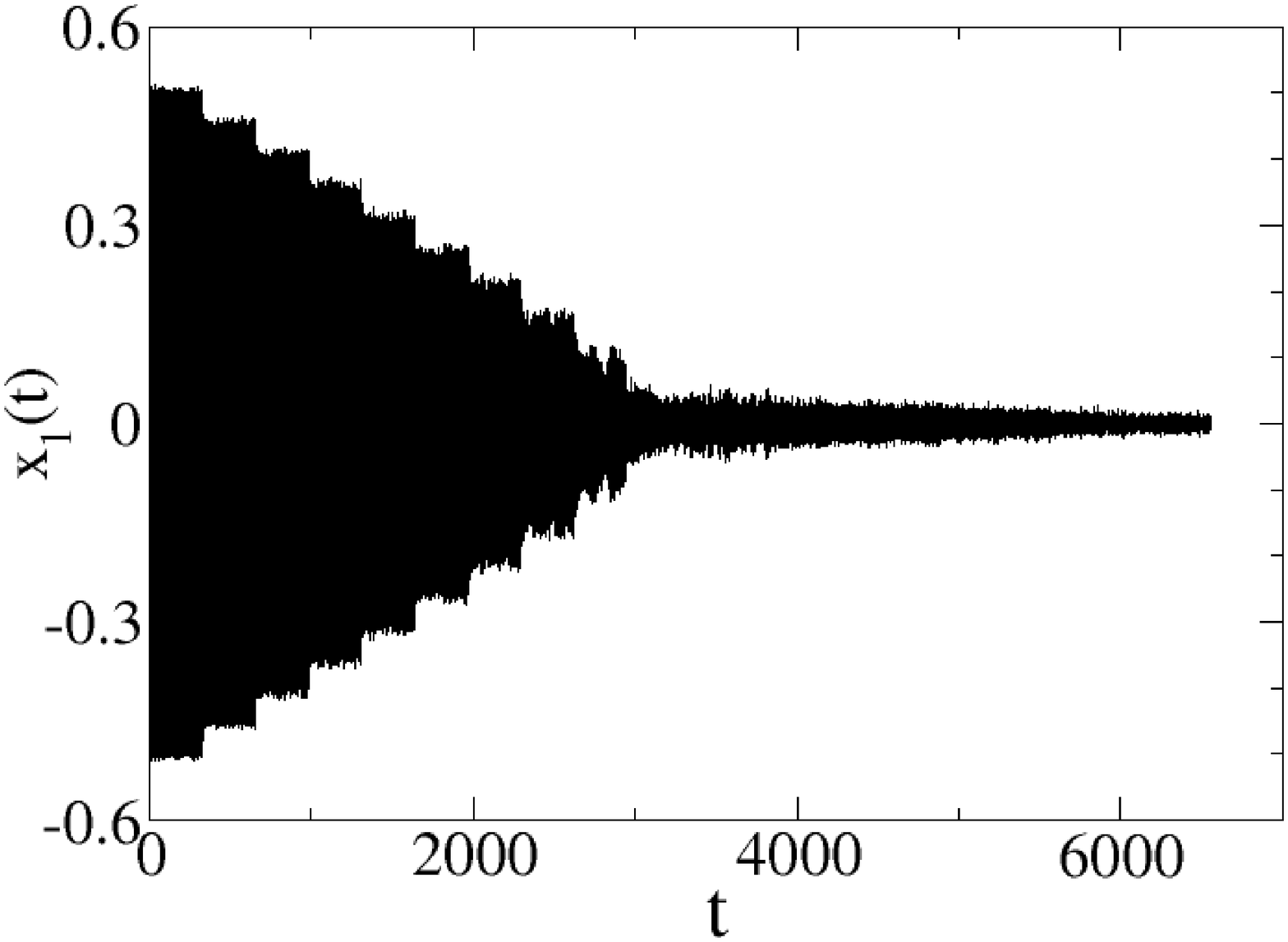}
\end{tabular}
\end{center}
\caption{\label{noisy_figs}Temporal behavior of noisy system
(\ref{Poinc:noise}) as $\epsilon$ is varied piecewisely.
Top: $\epsilon$ increases positively and OD is
observed near $\epsilon$=3.57. Bottom: $\epsilon$
decreases negatively and OD is observed near $\epsilon$=-1.8.
In both cases noise intensity is $D=1$.}
\end{figure}

The simulation of Eq.~(\ref{Poinc:noise}) is shown in
Fig.~\ref{noisy_figs}. The parameters used in this noisy
equation are the same as those used for Eq.~(\ref{Poinc:eq1}),
besides $D=1.0$. In the top panel, $\epsilon$ was piecewise
increased through a range of positive values, and OD was
observed to occur at a value very close to the noiseless case,
$\epsilon=3.57$. Similarly in Fig.~\ref{noisy_figs} (bottom)
$\epsilon$ was piecewise increased but now in the negative
direction. OD was still observed, at a value near that of the
noiseless case, $\epsilon=-1.8$. Thus the noise term in
Eq.~(\ref{Poinc:noise}) does not eliminate, or change the
nature of, the bifurcations leading to the appearance of OD;
however, it influences the asymptotic value of the fixed points
to which the system evolves.

 In our study of OD we have obtained analytic relations
predicting the onset of OD in the CV model, both for two and
five coupled oscillators, using assumptions provided by
experimental physiological measurements. In addition, all the
theoretical results have been verified in numerical
simulations. We have also observed {\it partial OD}
\cite{Liu:05} in our numerical model, where some oscillators
die while others remain active. It occurs if the eigenvalues
corresponding to certain variables $x_i$ and $y_i$ have
negative real parts, while those corresponding to other
oscillator variables do not have this property. For instance,
this happens if $2\alpha a_i + \epsilon < 0$. See
Fig.~\ref{x1_x2} (bottom).

Discussions of OD usually focus on the strength of the coupling
coefficient. The common phrase is that ``for large couplings,
amplitude death will take place". However, now that we have
analytic expressions for the critical value, also depending on
other parameters, we can predict the onset of OD when different
parameters are changed: OD is evidently a multi-parameter-sensitive
phenomenon. It can be induced, not only by changes in couplings,
but also by changes in the oscillator frequencies and amplitudes.
For instance, the analytic conditions for OD given by (\ref{condition:eq1})
indicate that, if the amplitude parameters $a_1$ and $a_2$
decrease (in fact it is sufficient to decrease the largest
one), then OD can occur for two coupled oscillators even when
the coupling coefficient remains constant. The same can happen
if the excitability parameter $\alpha$ is decreased below a
critical value.  Another interesting feature of OD inferred
from (\ref{condition:eq5}) and (\ref{five_algeb:eq4}) is that,
if $\epsilon > 0$, and $\epsilon > \epsilon_c$, new
asymptotically stable fixed points can appear. Thus the
oscillations die, but the dynamical variables can take non-zero
asymptotic values. We remark that the critical value
$\epsilon_c$ depends on the oscillators' frequencies. That is,
even for fixed coupling strength $\epsilon$, if the frequency
of one of the oscillators falls below some critical value, then
all the coupled system can stop dead. This behavior is true
both for two or five coupled oscillators.

What are the implications of these results for the CVS? Given
that (\ref{Poinc:eq1}) successfully models many features and
states of the CVS, we may speculate that phenomena analogous to
OD may also occur there. It would mean that, for certain
parameter combinations, the mutual interactions between the
cardiac, respiratory, myogenic, neurogenic and endothelial
oscillators might serve to bring one or all of them to a halt
-- or at least to inhibit function. Treatment to prevent or
remedy such a scenario might involve the use of drugs to modify
some of the parameters $\alpha$, $a_i$ or $\omega_i$ so as to
take the individual away from the regime of danger.

These results could also be applicable to coupled arrays of
neurons in the brain. In the first demonstration of OD in a
biological system Ozden et. al \cite{Ozden:04} coupled a
man-made device to an array of real neurons and OD was provoked
by taking the system into the strong-coupling regime. We
hypothesize that similar results could be obtained in response
to frequency changes, according to rules similar to
(\ref{condition:eq1}) or (\ref{condition:eq5}). The system
would not then need to be in the strong-coupling regime, which
in some cases could damage sensitive biological tissue.

In summary, we have shown analytically and through simulations that OD
in the CVS model (\ref{Poinc:eq1}) can occur via {\it either} of the
known bifurcation mechanisms. Furthermore, OD can be induced, not only
by an increase in coupling strength as conventionally accepted, but also
by changes of e.g.\ amplitude and/or frequency. We have shown it to be
sensitive to many different parameters. Connections between OD in the
model, and phenomena in the CVS, appear possible but remain to be
explored.
%%%%%%%%%%%%%%%%%%%%%%%%%  Acknowledgments  %%%%%%%%%%%%%%%%%%%%%%%%%%
\acknowledgments
This research was supported by the EU through the
\textbf{Al$\bm{\beta}$an} programme N\textsuperscript{o}
E03D05224VE, the FP6 BRACCIA programme, the Engineering and
Physical Sciences Research Council (UK), the Royal Society of
London, and the Wellcome Trust.
%%%%%%%%%%%%%%%%%%%%%%%%%  References  %%%%%%%%%%%%%%%%%%%%%%%%%%
% \bibliographystyle{apsrev}
% \bibliography{cardio21022007,chaosp}

\begin{thebibliography}{0}
\bibitem{foot1} The additional nitric oxide independent
    endothelial oscillatory process at about 0.007\,Hz
    \cite{Stefanovska:07} does not affect the
    results reported in this paper.
\end{thebibliography}

\end{document}